\PassOptionsToPackage{dvipsnames}{xcolor}
\documentclass[sigconf]{acmart}

\usepackage{booktabs} 

\copyrightyear{2017}
\acmYear{2017}
\setcopyright{acmcopyright}
\acmConference[ESPM2'17]{ESPM2'17: Third International Workshop on Extreme Scale Programming Models and Middleware}{November 12--17, 2017}{Denver, CO, USA}
\acmDOI{https://doi.org/10.1145/3152041.3152084}

\acmISBN{978-1-4503-5133-1/17/11}\acmPrice{\$15.00}

\usepackage{mdwtab}
\usepackage{notoccite}
\usepackage{comment}
\usepackage{color}
\usepackage[ampersand]{easylist}
\usepackage{booktabs}
\usepackage{listings}
\usepackage{bmpsize}
\usepackage{epstopdf}
\usepackage{array}
\usepackage{caption}
\usepackage{subcaption}
\usepackage{multicol}
\usepackage{fancyhdr}
\usepackage{lipsum}
\usepackage[english]{babel}
\usepackage{mathtools}
\usepackage[export]{adjustbox}
\usepackage{MnSymbol}
\usepackage{wasysym}
\usepackage{mdframed}
\newcolumntype{x}[1]{>{\centering\arraybackslash\hspace{0pt}}p{#1}}
\usepackage{array}
\usepackage[10pt]{moresize}
\newcolumntype{M}[1]{>{\centering\arraybackslash}m{#1}}
\newcolumntype{N}{@{}m{0pt}@{}}
\usepackage{balance}
\usepackage[export]{adjustbox}
\usepackage{epstopdf}
\usepackage{fixme}
\fxusetheme{color}
\fxsetup{
     status=draft,
     author=,
     layout=inline, 
     theme=color
}

\usepackage{multicol}
\usepackage{multirow}
\usepackage{comment}

\renewcommand{\footnoterule}{%
  \kern -3pt
  \hrule width \columnwidth height 0.1pt
  \kern 2pt
}
\definecolor{fxnote}{rgb}{0.8000,0.0000,0.0000}

\usepackage[T1]{fontenc}

\makeatletter
\def\@copyrightspace{\relax}
\makeatother

\renewcommand{\shortauthors}{Z. Khatami et al.}

\begin{document}
\vspace{-2cm}

\title{HPX Smart Executors}

\author{Zahra Khatami}
\affiliation{%
  \institution{\small Center for Computation and Technology\\
   Louisiana State University\\
       The STE$||$AR Group, http://stellar-group.org}
      \streetaddress{ Baton Rouge, LA, USA }
}
\email{zkhata@lsu.edu}

\author{Lukas Troska}
\affiliation{%
  \institution{\small Center for Computation and Technology\\
   Louisiana State University\\
       The STE$||$AR Group, http://stellar-group.org}
      \streetaddress{ Baton Rouge, LA, USA }
}
\email{lukas.troska@gmail.com}

\author{ Hartmut Kaiser}
\affiliation{%
  \institution{\small Center for Computation and Technology\\
   Louisiana State University\\
       The STE$||$AR Group, http://stellar-group.org}
      \streetaddress{ Baton Rouge, LA, USA }
}
\email{hkaiser@cct.lsu.edu}

\author{ J. Ramanujam}
\affiliation{%
  \institution{\small Center for Computation and Technology\\
   Louisiana State University\\
       The STE$||$AR Group, http://stellar-group.org}
      \streetaddress{ Baton Rouge, LA, USA }
}
\email{ ram@cct.lsu.edu}

\author{ Adrian Serio}
\affiliation{%
  \institution{\small Center for Computation and Technology\\
   Louisiana State University\\
       The STE$||$AR Group, http://stellar-group.org}
      \streetaddress{ Baton Rouge, LA, USA }
}
\email{aserio@cct.lsu.edu}

\renewcommand{\shortauthors}{Z. Khatami et al.}

\begin{abstract}
The performance of many parallel applications depends on loop-level parallelism. However, manually parallelizing all loops may result in degrading parallel performance, as some of them cannot scale desirably to a large number of threads. In addition, the overheads of manually tuning loop parameters might prevent an application from reaching its maximum parallel performance. We illustrate how machine learning techniques can be applied to address these challenges. In this research, we develop a framework that is able to automatically capture the static and dynamic information of a loop. Moreover, we advocate a novel method by introducing HPX smart executors for determining the execution policy, chunk size, and prefetching distance of an HPX loop to achieve higher possible performance by feeding static information captured during compilation and runtime-based dynamic information to our learning model. Our evaluated execution results show that using these smart executors can speed up the HPX execution process by around $12\%-35\%$ for the Matrix Multiplication, Stream and $2D$ Stencil benchmarks compared to setting their HPX loop's execution policy/parameters manually or using HPX auto-parallelization techniques.
\end{abstract}


\maketitle

\section{Introduction}

Runtime information is often speculative. While runtime adaptive methods have been shown to be very effective -- especially for highly dynamic scenarios -- solely relying on them doesn't guarantee maximal parallel performance, since the performance of an application depends on both the values measured at runtime and the related transformations performed at compile time. Collecting the outcome of the static analysis performed by the compiler could significantly improve runtime decisions and therefore application performance \cite{ML7,ML2,z1,z2}.

There are many existing publications on automatically choosing optimization parameters based on static information extracted at compile time. For example in \cite{run1,run2} optimal scheduling for parallel loops is implemented dynamically at runtime by examining data dependencies captured at compile time. However, one of the challenges in these studies is the need to repeat their proposed methods for each new program, which in general is not desirable, as it requires extra execution time for each application for such parameters determination. Moreover, manually tuning parameters becomes ineffective and almost impossible when the parallel performance depends on too many parameters as defined by to the program. 
Hence, many researches have extensively studied machine learning algorithms which optimize such parameters automatically.

For example in \cite{ML6}, a neural network and decision tree are applied on the training data collected from different observations to predict the branch behavior in a new program. In \cite{ML7} nearest neighbors and support vector machines are used for predicting unroll factors for different nested loops based on the extracted static features. In \cite{ML11,ML8}, the logistic regression model is used to derive a learning model, which results in a significant speedup in compilation time of their studied benchmarks. Most of these optimization techniques require users to compile their application twice, first compilation for extracting static information and the second one for recompiling application based on those extracted data. None of these considers both static and dynamic information. 

The goal of this paper is to optimize an HPX application's performance by predicting optimum parameters for its parallel algorithms by considering both static and dynamic information and to avoid unnecessary compilation. As all of the HPX parallel algorithms perform based on the dynamic analysis provided at runtime, this technique is unable to achieve the maximum possible parallel efficiency in some cases: 

\begin{itemize}
\item In \cite{hpx3,hpx4} different policies for executing HPX parallel algorithms are studied. However these policies should be manually selected for each algorithm within an application, which may not be an optimum approach, as a user should execute each parallel algorithm of his application with different execution policies to find the efficient one for that algorithm. 

\item Determining chunk size is another challenge in the existing version of the HPX algorithms. Chunk size is the amount of work performed by each task \cite{chunk1,hpx5} that is determined by an \textit{auto\_partitioner} exposed by the HPX algorithms or is passed by using \textit{static/dynamic\_chunk\_size} as an execution policy's parameter \cite{hpx3}. However,
\begin{enumerate}
\item the experimental results in \cite{z2} and \cite{z1} showed that the overheads of determining chunk size by using the \textit{auto\_partitioner} negatively effected the application's scalability in some cases;  
\item the policy written by the user will often not be able to determine the optimum chunk size either due to the limit of runtime information. 
\end{enumerate}

\item In \cite{z3}, we proposed the HPX prefetching method which aids prefetching that not only reduces the memory accesses latency, but also relaxes the global barrier. Although it results in better parallel performance for the HPX algorithms, however, a distance between each two prefetching operations should also be manually chosen by a user for each new program. 
\end{itemize}

Automating these mentioned parameters selections by considering loops characteristics implemented in a learning model can optimize the HPX parallel applications performances. To the best of our knowledge, we present the first attempt to implement a learning model for predicting optimum loop parameters at runtime, wherein the learning model captures features both from static compile time information and from runtime introspection. 

In this research, we introduce a new ClangTool \textit{ForEachCallHandler} using LibTooling \cite{libtooling} as a custom compiler pass to be executed by the Clang compiler, which is intended to collect the static features at compile time. The logistic regression model is implemented in this paper as a learning model that considers these captured features for predicting efficient parameters for an HPX loop. For implementing this learning model on a loop, we propose new HPX smart executors that -- when used on a parallel loop -- instructs the compiler to apply this \textit{ForEachCallHandler} tool on that loop. As a results, the loop's features will automatically be included in the prediction process implemented with that learning model. One of the advantages of this approach in utilizing HPX policies is that in practice it enables us to change the algorithms internal structure at runtime and therefore we do not have to compile the code again after the code transformation step. 

This technique is able to use high-level programming abstractions and machine learning to relieve the programmer of difficult and tedious decisions that can significantly affect program behavior and performance. Our results show that combining machine learning, compiler optimizations and runtime adaptation helps us to maximally utilize available resources. This improves application performance by around $12\%-35\%$ for the Matrix Multiplication, Stream and $2D$ Stencil benchmarks compared to setting their HPX loop's execution policy/parameters manually or using HPX auto-parallelization techniques. 

The remainder of this paper is structured as follows: the machine learning algorithms that are used to study the learning models are discussed in section \ref{sec:learnAlg}; the proposed model is discussed is more details in section \ref{sec:model}, and section \ref{sec:res} provides the experimental results of this proposed technique. Conclusion and future works are explained in section \ref{sec:con}.

\section{Learning Algorithm}
\label{sec:learnAlg}

In this research we use the binary and multinomial logistic regression models~\cite{bishop1} to select the optimum execution policy, chunk size, and prefetching distance for certain HPX loops based on both, static and dynamic information, with the goal of minimizing execution time. Logistic regression model has been used in several previous works \cite{ML11, MLa}, and it is shown to be able to predict such parameters accurately. We will show later that the performance of these learning models has high accuracy for about $98\%$ and $95\%$ for the binary and multinomial logistic regression models respectively on the studied problems. Also, compared to the other learning models such as artificial neural networks (ANNs), the implemented logistic regression model has lower computational complexity. Moreover, since the chunk size values can be seen as a categorical variables, this makes the logistic regression models well-suited for our problem. 

Here, the static information about the loop body (such as the number of operations, see Table~\ref{table:t2}) collected by the compiler and the dynamic information (such as the number of cores used to execute the loop) as provided by the runtime system is used to feed a logistic regression model enabling a runtime decision to obtain highest possible performance of the loop under consideration. The presented method relies on a compiler-based source-to-source transformation. The compiler transforms certain loops which were annotated by the user by providing special executors -- discussed later in section \ref{execp} -- into code controlling runtime behavior. This transformed code instructs the runtime system to apply a logistic regression model and to select either an appropriate code path (e.g. parallel or sequential loop execution) or certain parameters for the loop execution itself (e.g. chunk size or prefetching distance). We briefly discuss these learning models in the following sections.

\subsection{Binary Logistic Regression Model }
\label{sec:bin}

In order to select the optimum execution policy (sequential or parallel) for a loop, the binary logistic regression model is implemented to analyze the static information extracted from the loop by the compiler and the dynamic information as provided by the runtime. In this model, the weights parameters having $k$ features $W^T = [\omega _1, \omega _2, ..., \omega _{k}]$ are determined by considering features values $x_r (i)$ of each experiment $X_i = [1, x_1 (i), ..., x_{k} (i)]^T$ which minimize the log-likelihood of the Bernoulli distribution value as follow:
\begin{equation} 
\mu_i = 1/(1 + e^{-W^T X_i}). 
\end{equation}
 
The values of $\omega$ are updated in each step $t$ as follows:
\begin{equation} \label{eq1}
\omega_{t + 1} = (X^TS_tX)^{-1}X^T(S_tX\omega_t + y - \mu_t)
\end{equation}

\noindent
, where $S$ is a diagonal matrix with $S(i, i) = \mu_i(1 - \mu_i)$. The output is determined by considering the following decision rule:
\begin{equation} 
\label{eq2}
y(x) = 1 \longleftrightarrow p(y = 1 | x) > 0.5
\end{equation}

\subsection{Multinomial Logistic Regression Model }
\label{sec:mult}

In order to predict the optimum values for the chunk size and the prefetching distance, the multinomial logistic regression model is implemented to analyze the static information extracted from the loop by the compiler and the dynamic information as provided by the runtime. If we have $N$ experiments that are classified in $C$ classes and each has $K$ features, the posterior probabilities are computed by using a softmax transformation of the features variables linear functions for an experiment $n$ with a class $c$ as follow:

\begin{equation}
\label{eq3}
y_{nc} = y_c(X_n) = \frac{exp(W_c^TX_n)}{\sum_{i=1}^{C} exp(W_i^TX_n)}
\end{equation}

\noindent
The cross entropy error function is defined as follows:

\begin{equation}
\label{eq4}
E(\omega_1, \omega_2, ..., \omega_C) = -\sum_{n=1}^{N}\sum_{c=1}^{C} t_{nc} lny_{nc}
\end{equation}

\noindent
, where T is a $N\times C$ matrix of target variables with $t_{nc}$ elements. The gradient of $E$ is computed as follows:

\begin{equation}
\label{eq5}
\nabla_{\omega_{c}}E(\omega_1, \omega_2, ..., \omega_C) = \sum_{n=1}^{N}(y_{nc} - t_{nc})X_n
\end{equation}

In this method, we use the Newton-Raphson method \cite{nr} to update the weights values in each step:

\begin{equation}
\label{eq6}
\omega_{new} = \omega_{old} - H^{-1}\nabla E(\omega)
\end{equation}

\noindent
where $H$ is the Hessian matrix defined as follows:

\begin{equation}
\label{eq7}
\nabla_{\omega_{i}}\nabla_{\omega_{j}} E(\omega_1, \omega_2, ..., \omega_C)=\sum_{n=1}^{N} y_{ni}(I_{ij} - y_{nj})X_n X^T_n 
\end{equation}

\noindent
More details can be found in \cite{bishop2}.  

\section{Proposed Model}
\label{sec:model}

In this section, we propose a new technique for applying the learning models discussed in section \ref{sec:learnAlg} to HPX loops. The goal of this technique is to combine machine learning methods, compiler transformations, and runtime introspection in order to maximize the use of available resources and to minimize execution time of the loops. Its design and implementation has several steps categorized as follow\footnotemark[1]: 

\begin{enumerate}
\item New HPX Smart Executors
\item Features Extraction
\item Design of Learning Model
\item Learning Model Implementation
\end{enumerate}

\footnotetext[1]{This technique with its installation instructions are publicly available at \url{https://github.com/STEllAR-GROUP/hpxML}. Feel free to join our IRC channel $\#ste||ar$ if you need any help.}

\subsection{New HPX Smart Executors}
\label{execp}

We introduce two new HPX execution policies and one new HPX execution policy parameter, which we refer to them as the \textit{smart} executors in this paper, since they enable the weights gathered by the learning model to be applied on the loop. \textit{par\_if} and\\ \textit{make\_prefetcher\_policy} as the smart policies instrument executors to be able to consume the weights produced by a binary logistic regression model, which is used to select the execution policy corresponding to the optimal code path to execute (sequential or parallel), and a multinomial logistic regression model, which is used to determine an efficient prefetching distance. \textit{adaptive\_chunk\_size} as the smart execution policy parameter uses a multinomial logistic regression model to determine an efficient chunk size. Fig.\ref{fa} shows three loops defined with these smart execution policies and parameter that apply a \textit{lambda} function over a \textit{range}. We have created a new special compiler pass for clang which recognizes these annotated loops and transform them into equivalent code which instructs the runtime to apply the described regression models.

\begin{figure}
    \begin{lstlisting}[basicstyle=\scriptsize] 
for_each(par_if, range.begin(), range.end(), lambda); 

for_each(policy.with(adaptive_chunk_size()),range.begin(),range.end(),lambda); 

for_each(make_prefetcher_policy(policy,
          prefetching_distance_factor, 
          container_1,..., container_n),
          range.begin(),range.end(),lambda); 	
    \end{lstlisting}\vspace{-0.15cm}
    \caption{\small{Loops using the proposed smart executors, which are recognized and instrumented by the compiler to allow HPX to consider the weights produced by the learning models when executing the loops.}}
    \label{fa}
\end{figure}

\begin{table}
\centering
    \begin{tabular}{|c|c|}
       \hline
       static/dynamic & Information\\
       \hline
        dynamic & {\color{red} number of threads$*$}\\
        \hline
       dynamic & {\color{red} number of iterations$*$}\\
        \hline
        static & {\color{red} number of total operations per iteration$*$}\\
        \hline
        static & {\color{red} number of float operations per iteration$*$}\\
        \hline
        static & {\color{red} number of comparison operations per iteration$*$}\\
        \hline
        static & {\color{red} deepest loop level$*$}\\
        \hline
        static & number of integer variables\\
        \hline
        static & number of float variables\\
        \hline
        static & number of if statements\\
        \hline
        static & number of if statements within inner loops\\
        \hline
        static & number of function calls\\
        \hline
        static & number of function calls within inner loops\\
    \hline
    \end{tabular}
    \caption{\small{Collected static and dynamic features. First $6$ features marked with {\color{red} red*} have been selected for our model using the decision tree classification technique \cite{dt1, dt2} to avoid overfitting the model.}}
    \label{table:t2}
\end{table}

\subsection{Features Extraction}

Initially, we selected $10$ static features to be collected at compile time and $2$ dynamic features to be determined at runtime to be evaluated by our learning model. These features are listed in Table \ref{table:t2}. Although it may not be the best possible set, it is very similar to those considered in the other works \cite{ML7, ML2, ML11}, which in their results indicated that the set is sufficient to design a learning model for this type of problem. To avoid overfitting the model, we chose $6$ critical features marked with {\color{red}red$*$} color in Table \ref{table:t2} to include in the actual decision tree classification technique\cite{dt1, dt2}, which reduces the initial features set in a tree building process based on information gain value. This value is used to decide which feature to be selected for splitting data at each step in a tree building process. More information about this technique can be found in \cite{dt3, dt4}.

In order to collect static information at compile time,  we introduce a new ClangTool named \textit{ForEachCallHandler} in the Clang compiler as shown in Fig.\ref{f1}. This tool locates in the user source code instances of loops which use the proposed smart executors. Once identified, the loop body is then extracted from the $lambda$ function by applying \textit{getBody()} on a $lambda$ operator \textit{getLambdaCallOperator()}. The value of each of the listed static features is then recorded by passing $lambda$ to \textit{analyze\_statement}. In order to capture dynamic features at runtime, the compiler inserts hooks (HPX API function calls) which are invoked by the runtime. In this instance the compiler will insert the call \textit{hpx::get\_os\_thread\_count()} and \textit{std::distance(range.begin(), range.end())} which will return the number of OS threads as well as the number of iterations that the loop will run over, respectively.

\begin{figure}
    \begin{lstlisting}[basicstyle=\scriptsize] 
class ForEachCallHandler: 
     public RecursiveASTVisitor<ForEachCallHandler> { ...  
	//Visit every call expression
	bool VisitCallExpr(const CallExpr *call) { ...
		//check if a call is an HPX algorithm
		if (func_string.find("hpx::parallel") !=string::npos) {
		
		  //Capturing lambda function from a loop
		  const CXXMethodDecl* lambda_callop =
	           lambda_record->getLambdaCallOperator();
		  Stmt* lambda_body = lambda_callop->getBody();
		
		  //Capturing policy
	 	  SourceRange policy(call->getArg(0)->getExprLoc(), 
	 	   call->getArg(1)->getExprLoc().getLocWithOffset(-2));
		   
		  //Extracting static/dynamic features
		  analyze_statement(lambda_body);
                                      
		  //Determining policy
	 	  if (policy_string.find("par_if") != string::npos)
	 	     policy_determination(call, SM); 
	 	
		  //Determining chunk size
	 	  if(policy_string.find("adaptive_chunk_size")!=string::npos)
	 	     chunk_size_determination(call, SM);
		
		  //Determining prefetching distance
	 	  if(policy_string.find("make_prefetcher_policy")!=string::npos)
	 	     prefetching_distance_determination(call, SM);
		     ...
}}}
    \end{lstlisting}
    \captionsetup{singlelinecheck=off}\vspace{-0.15cm}
    \caption[]{\small{The proposed ClangTool \textit{ForEachCallHandler} to collecting static/dynamic information of each loop and implement a learning model based on 
                             the current smart executors. }}
    \label{f1}
\end{figure}

\subsection{Designing the Learning Model}
\label{design}

To design an efficient learning model that could be able to cover various cases, we collected over $300$ training data sets by analyzing Matrix multiplication application with different problem sizes that implements \textit{par\_if}, \textit{adaptive\_chunk\_size} or \textit{make\_prefetcher\_policy} on its loops. The experimental results evaluated in Section \ref{sec:res} show that these training data\footnotemark[1] are enough to predict the HPX loop's parameters accurately for the studied applications: Matrix multiplication, Stream and $2D$ Stencil benchmarks. The regression models are designed based on these collected data, in which the values of $\omega$ from eq.\ref{eq1} and eq.\ref{eq6} are determined whenever the sum of square errors reaches its minimum value. Then they are stored in an output file named as \textit{weights.dat} that will be used for predicting the optimal execution policy, chunk size, and prefetching distance at runtime. This learning step can be done offline, which also doesn't add any overhead at compile time nor does at runtime.

\footnotetext[1]{The characteristics of the loops of these training data are available at \url{https://github.com/STEllAR-GROUP/hpxML/blob/master/logisticRegressionModel/algorithms/inputs}.}

It should be noted that the multinomial logistic regression model must be initialized with the allowed boundaries for the chunk size and prefetching distance in order to choose an efficient value. In this study we selected $0.1\%$, $1\%$, $10\%$, or $50\%$ of the iterations of a loop as chunk size candidates and $1$, $5$, $10$, $100$ and $500$ cache lines as prefetching distance candidates. These candidates are validated with different tests and based on their results, they are selected. In order to derive the fidelity of the model, we train the algorithm using $80\%$ of the test cases and use the remaining $20\%$ as a trials to see how accurate the predictions are. Our results show that the binary logistic regression model is accurate in $98\%$ of the trials and the multinomial logistic regression model is accurate in $95\%$ of them.

\subsection{Learning Model Implementation}
\label{sublearn}

\subsubsection{Binary Logistic Regression Model (Execution Policy)}

We propose a new function \textit{seq\_par} that passes the extracted features for a loop that uses \textit{par\_if} as its execution policy. In this technique, a Clang compiler automatically adds extra lines within a user's code as shown in Fig.\ref{f2} that allows the runtime system to decide whether execute a loop sequentially or in parallel based on the return value of \textit{seq\_par} from Eq.\ref{eq2}. If the output is $false$ the loop will execute sequentially and if the output is $true$ the loop will execute in parallel.

\begin{figure}
\begin{subfigure}{0.5\textwidth}
    \begin{lstlisting}[basicstyle=\scriptsize]  
for_each(par_if,range.begin(),range.end(),lambda); 		  
    \end{lstlisting}\vspace{-0.15cm}
    \caption{\small{Before compilation}}
    \label{f2a}
    \end{subfigure}
     
    \begin{subfigure}{0.5\textwidth}
    \begin{lstlisting}[basicstyle=\scriptsize]                           
if(seq_par({f0,..fn})) //extracted static information
	for_each(seq, range.begin(),range.end(),lambda);  
else
	for_each(par, range.begin(),range.end(),lambda); 		  
    \end{lstlisting}\vspace{-0.15cm}
    \caption{\small{After compilation}}
    \label{f2b}
    \end{subfigure}\vspace{-0.15cm}
     \caption{\small{The proposed function \textit{seq\_par} for a implementing binary logistic regression model at runtime.}}
    \label{f2}
\end{figure}

\subsubsection{Multinomial Logistic Regression Model (Chunk Size)}
\label{ch}

We propose a new function \textit{chunk\_size\_determination} that passes the extracted features for a loop that uses \textit{adaptive\_chunk\_size} as its execution policy's parameter. In this technique, a Clang compiler changes a user's code automatically as shown in Fig.\ref{f3} that makes runtime system to choose an optimum chunk size based on the output of \textit{chunk\_size\_determination} from Eq.\ref{eq3} that is based on the chunk size candidate's probability and it is computed using the values of the studied loop's features and the learning model's weights.

\begin{figure}
\begin{subfigure}{0.5\textwidth}
    \begin{lstlisting}[basicstyle=\scriptsize]  
for_each(policy.with(adaptive_chunk_size()),
         range.begin(),range.end(),lambda); 		  
    \end{lstlisting}\vspace{-0.15cm}
    \caption{\small{Before compilation}}
    \label{f3a}
    \end{subfigure}
    \begin{subfigure}{0.5\textwidth}
    \begin{lstlisting}[basicstyle=\scriptsize]
for_each(policy.with(chunk_size_determination({f0,...fn}))), 
         range.begin(),range.end(),lambda);		  
    \end{lstlisting}\vspace{-0.15cm}
    \caption{\small{After compilation}}
    \label{f3b}
    \end{subfigure}\vspace{-0.15cm}
     \caption{\small{The proposed function \textit{chunk\_size\_determination} for implementing a multinomial logistic regression model at runtime.}}
    \label{f3}
\end{figure}

\subsubsection{Multinomial Logistic Regression Model (Prefetching Distance)}

We propose a new function \textit{prefetching\_distance\_determination} that passes the extracted features for a loop that uses \textit{make\_prefetcher\_policy} as its execution policy. In this technique, a Clang compiler changes a user's code automatically as shown in Fig.\ref{f4} that makes runtime system to choose an optimum prefetching distance based on the output of \textit{prefetching\_distance\_determination}. This function also computes the outputs by implementing Eq.\ref{eq3} using the values of the studied loop's features and the learning model's weights.

\begin{figure}
\begin{subfigure}{0.5\textwidth}
    \begin{lstlisting}[basicstyle=\scriptsize]  
for_each(make_prefetcher_policy(policy,
          prefetching_distance_factor, 
          container_1,..., container_n),
          range.begin(),range.end(),lambda); 		  
    \end{lstlisting}\vspace{-0.15cm}
    \caption{\small{Before compilation}}
    \label{f4a}
    \end{subfigure}
    \begin{subfigure}{0.5\textwidth}
    \begin{lstlisting}[basicstyle=\scriptsize]
for_each(make_prefetcher_policy(policy,
          prefetching_distance_determination({f0,...fn}), 
          container_1,..., container_n),
          range.begin(),range.end(),lambda); 	  
    \end{lstlisting}\vspace{-0.15cm}
    \caption{\small{After compilation}}
    \label{f4b}
    \end{subfigure}\vspace{-0.15cm}
     \caption{\small{The proposed function \textit{prefetching\_distance\_determination} for implementing a multinomial logistic regression model at runtime.}}
    \label{f4}
\end{figure}

As we can see, these proposed techniques consider both, the static and the dynamic information for determining an efficient execution policy, chunk size, and prefetching distance for a loop. In addition, this decision process is performed at runtime by computing outputs of \textit{seq\_par}, \textit{chunk\_size\_determination} and \textit{prefetching\_distance\_determination}, which avoids an extra compilation step. In other words, static information is collected during compilation and the decisions aiming at optimum parameters are made at runtime while taking into account additional runtime information. One of the other advantages of this method are that other parameters and executors attached to the current executors can be also reattached to the generated execution policy. Moreover, all of these smart executors can be used together by simply defining a loop policy to be ``\textit{make\_prefetcher\_policy(par\_if, ...).with(adaptive\_chunk\_size())}". The experimental results of our proposed learning techniques discussed are presented in the next section.

\section{Experimental Results}
\label{sec:res}

In this section, we evaluate the performance of our proposed technique using Clang $4.0.0$ and HPX $V0.9.99$ on the test machine with two Intel Xeon E5-2630 processors, each with $8$ cores clocked at $2.4GHZ$ and $65GB$ of main memory. The main goal here is to illustrate that dynamic information obtained at runtime and static information obtained at compile time are both necessary to provide sufficient parallel performance and the proposed techniques are able to predict the optimum parameters for HPX loops based on these information\footnotemark[1].

\footnotetext[1]{Applications evaluated in this Section are publicly available at \url{https://github.com/STEllAR-GROUP/hpxML/tree/master/examples}.}

\subsection{Artificial Test Cases}

\begin{table*}
\centering
    \begin{tabular}{|c|c|c|c|c|c|c|c|c|c|}
       \hline
      Test &Loop & Iterations & Total opr. & Float opr. & Comparison opr. & Loop level & Policy (Threads) & Chunk size\% & Pref. dist.\\
       \specialrule{0.2em}{0.01em}{0.01em}
      \multirow{2}{*}{1} & $l_1$ & 10000 & 400100 & 200000 & 101010 & 2 & par (8) & 0.1 & 5\\
       & $l_2$ & 20000 & 450026 & 250000 & 150503 & 2 & par (8) & 0.1 & 5\\
       & $l_3$ & 20000 & 502040 & 250000 & 103051 & 2 & par (8) & 0.1 & 1\\
       & $l_4$ & 500 & 550402 & 200000 & 150102 & 1 & par (8) & 10 & 5\\
      \specialrule{0.2em}{0.01em}{0.01em}
       \multirow{2}{*}{2} & $l_1$ & 150000 & 350106 & 101010 & 500 & 2 & par (8) & 0.1 & 10\\
       & $l_2$ & 100 & 10050016 & 5000000 & 2505013 & 3 & seq & 10 & 1\\
       & $l_3$ & 100 & 25000000 & 3010204 & 1500204 & 3 & seq & 10 & 1\\
       & $l_4$ & 50000 & 4000450 & 200000 & 100150 & 1 & par (8) & 1 & 5\\
       \specialrule{0.2em}{0.01em}{0.01em}
       \multirow{2}{*}{3} & $l_1$ & 500 & 4504030 & 250000 & 150300 & 2 & par (8) & 1 & 10\\
       & $l_2$ & 400 & 3502020 & 200000 & 100405 & 1 & par (8) & 1 & 10\\
       & $l_3$ & 2000 & 250033 & 150000 & 103040 & 3 & seq & 10 & 5\\
       & $l_4$ & 2500 & 350400 & 150000 & 100600 & 3 & seq & 10 & 5\\
       \specialrule{0.2em}{0.01em}{0.01em}
       \multirow{2}{*}{4} & $l_1$ & 20000 & 204002 & 100000 & 10320 & 2 & par (8) & 0.1 & 1\\
       & $l_2$ & 30000 & 400000 & 150102 & 10000 & 2 & par (8) & 0.1 & 1\\
       & $l_3$ & 300 & 550000 & 44000 & 20030 & 3 & seq & 10 & 5\\
       & $l_4$ & 400 & 450000 & 50400 & 10602 & 3 & seq & 10 & 10\\
       \specialrule{0.2em}{0.01em}{0.01em}
       \multirow{2}{*}{5} & $l_1$ & 200 & 4502001 & 150000 & 101004 & 3 & par (8) & 1 & 1\\
       & $l_2$ & 700 & 400020 & 300000 & 150006 & 3 & par (8) & 1 & 5\\
       & $l_3$ & 300 & 302020 & 20000 & 14005 & 2 & par (8) & 1 & 5\\
       & $l_4$ & 100 & 50400 & 20000 & 10110 & 2 & seq & 10 & 10\\
       \hline
    \end{tabular}
    \caption{\small{Execution policy, chunk size and prefetching distance determined by the proposed techniques based on the static/dynamic information extracted from each loop and the weights provided by the learning models.}}
    \label{t1}
\end{table*}

In this section, we evaluate the performance of the proposed techniques from Section \ref{sec:model} over $5$ different artificial test cases shown in Table \ref{t1}, in which each of them includes $4$ loops with different characteristics. Each of these loops of each test cases is a Matrix multiplication computation with different problem sizes included in this table. The main purpose of these evaluations is to show the effectiveness of each proposed method on an HPX parallel performance. 

\subsubsection{\textit{par\_if}}

Parallelizing all loops within an application may not result in a best possible parallelization, as some of the loops cannot scale desirably on more number of threads. For evaluating the effectiveness of the proposed \textit{seq\_par} function exposed by a smart executor \textit{par\_if} discussed in section \ref{sec:model}, we study its implementation on the described $5$ test cases. These test cases are selected to show that in case of having several loops within a parallel application, some of these loops should be executed in sequential to achieve a better parallel performance. Each of these test cases is executed three times by setting execution policies of the outer loops to be \textit{seq}, \textit{par}, or \textit{par\_if} in each time. The static and dynamic characteristics of each loop in each test are listed in Table \ref{t1}. The execution policies determined by using \textit{par\_if} policy for each loop are also included in the column \textit{Policy} of this Table.

\begin{figure}
\begin{center}
\centering
\includegraphics[width=0.9\columnwidth]{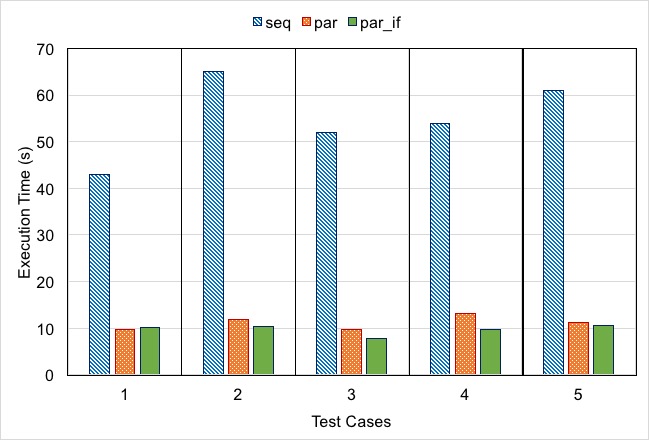}\vspace{-0.15cm}
\caption {\small{The execution time comparisons between setting execution policy of the loops to be 
                           \textit{seq}, \textit{par}, or \textit{par\_if}.}}
\label{f6}
\end{center}
\end{figure}

Fig.\ref{f6} shows the execution time for each test case and it illustrates that in most of them using \textit{par\_if} will outperform the basic policy \textit{par}. The main reason of this improvement is that by considering the determined execution policy included in Table \ref{t1}, as the execution policy \textit{seq} is determined for some of the loops that cannot scale desirably on more number of threads, this technique results in outperforming manually parallelized code by around $15\% - 20\%$ for these test cases expect the first one. In this test case, however, the total execution time of the loops took slightly longer when invoked with \textit{par\_if}. This is due to the overhead generated during the invocation of the binary logistic regression model's cost function, manually setting their execution policy as \textit{par} resulted in having a better performance.

\subsubsection{\textit{adaptive\_chunk\_size}}

As discussed in Section \ref{sec:model}, the proposed \textit{chunk\_size\_determination} function exposed by a smart executor \textit{adaptive\_chunk\_size} enables the runtime system to choose an efficient chunk size for a loop by considering static and dynamic features of that loop. As mentioned in section \ref{design}, this method selects between chunk sizes of $0.1\%$, $1\%$, $10\%$, or $50\%$ of the iterations of a loop by comparing their probabilities in the multinomial logistic regression model's cost function.

\begin{figure}
\begin{center}
\centering
\includegraphics[width=0.9\columnwidth]{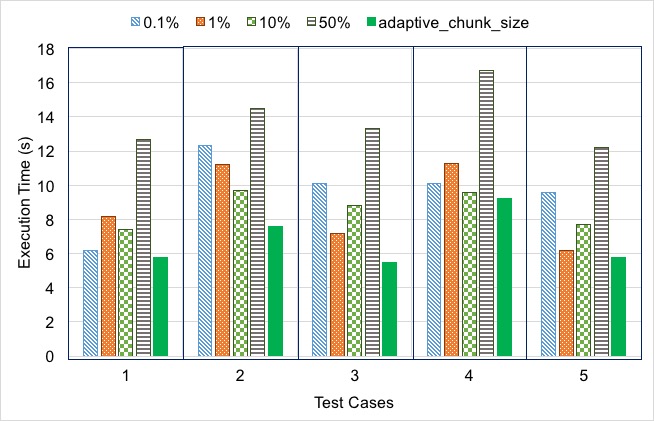}\vspace{-0.15cm}
\caption {\small{The execution time comparisons between setting chunk size of the loops to be $0.1\%$, $1\%$, $10\%$, or $50\%$ of the iterations of a loop and the chunk size determined by using \textit{adaptive\_chunk\_size}.}}
\label{f7}
\end{center}
\end{figure}

Fig.\ref{f7} shows the execution time for each test case in Table \ref{t1} by setting optimal chunk size of each loop. The chunk size determined by the algorithm for each loop are also included in the column \textit{Chunk size\%} of the Table \ref{t1}. The overall performance of these cases show by an average of about $31\%$, $15\%$, $17\%$ and $38\%$ improvement over setting chunks to be $0.1\%$, $1\%$, $10\%$, or $50\%$ of the iterations of a loop. The main reason of this improvement is that efficient chunk size helps in having even amount of work on each number of threads that results in reducing total overheads and latencies. These results also illustrate the importance of the chunk size's effect on an application's scalability and the capability of this method in improving parallel performance of an application by choosing efficient chunk size for each loop.

\subsubsection{\textit{make\_prefetcher\_policy}}

As discussed in Section \ref{sec:model}, the proposed \textit{perfecting\_distance\_determination} function exposed by a smart executor \textit{make\_prefetcher\_policy} allows the runtime system to choose an efficient prefetching distance for a loop by considering static and dynamic features of that loop. As it mentioned in Section \ref{design}, this method chooses between prefetching distances of $1$, $5$, $10$, $100$ and $500$ cache lines by comparing their probabilities in the multinomial logistic regression model's cost function. 

\begin{figure}
\begin{center}
\centering
\includegraphics[width=0.9\columnwidth]{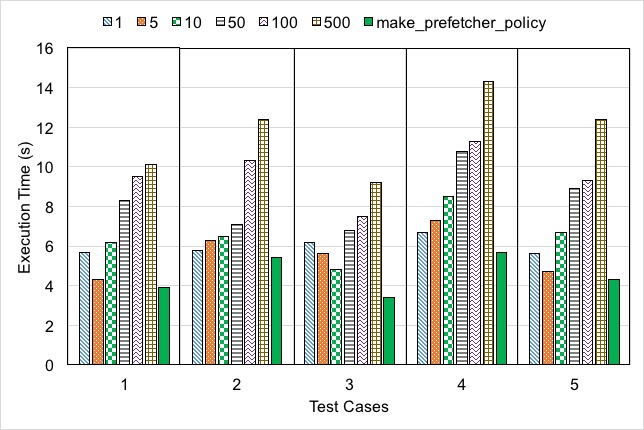}\vspace{-0.15cm}
\caption {\small{The execution time comparisons between setting the prefetching distance of the 
                           loops to be $1$, $5$, $10$, $100$, or $500$ cache lines and the determined 
                           prefetching distance using \textit{make\_prefetcher\_policy}.}}
\label{f8}
\end{center}
\end{figure}

Fig.\ref{f8} shows the execution time for each prefetching size in each test case in Table \ref{t1}. The prefetching distance determined by the algorithm for each loop are also included in the last column of the Table \ref{t1}. The overall performance of these cases show by an average of about $25\%$, $19\%$, $14\%$, $33\%$, $24\%$, and $47\%$ improvement over setting prefetching distances to be $1$, $5$, $10$, $100$, or $500$ cache lines. The main reason of this improvement is that using efficient prefetching distance resulted in better cache usage that reduced the total overheads. 

\subsection{Real Benchmarks}

In the previous section, we demonstrated the effectiveness of each proposed on an HPX parallel performance on $5$ different test cases in which each of them includes $4$ different loops for a matrix multiplication computation. 
In this section, we apply all of the proposed methods together on two different benchmarks: the Stream and $2D$ Stencil benchmarks. 
The previous training data is also used in the proposed techniques applied on these applications.

\begin{table}
\footnotesize
\centering
    \begin{tabular}{|c|c|c|c|c|c|}
       \hline
      Test & Iterations & Total opr. & Float opr. & Comp. opr. & level\\
       \specialrule{0.1em}{0.01em}{0.01em}
      \multirow{1}{*}{Stream} & 50000000 & 8 & 8 & 0 & 0\\
      \specialrule{0.1em}{0.01em}{0.01em}
       \multirow{1}{*}{Stencil} & 45 & 3502 & 2500 & 301 & 1\\
       \hline
    \end{tabular}
    \caption{\small{Dynamic/static features for each benchmark.}}
    \label{t3}
\end{table}

\subsubsection{Stream Benchmark}

\begin{figure}
    \begin{lstlisting}[basicstyle=\scriptsize]
for_each(policy, a_begin, a_end,[&](std::size_t i){
                 c[i] = a[i]; // copy step
                 b[i] = k * c[i]; // scale step
                 c[i] = a[i] + b[i]; // adding step
                 a[i] = b[i] + k * c[i]; // triad step
              });
    \end{lstlisting}\vspace{-0.15cm}
    \caption{\small{Stream Benchmark.}}
    \label{f14}
\end{figure}

\begin{figure}
\begin{center}
\centering
\includegraphics[width=0.6\columnwidth]{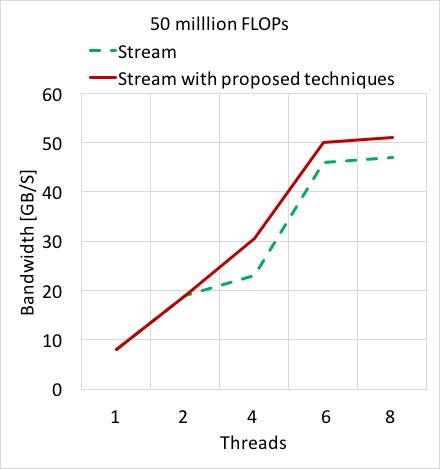}\vspace{-0.15cm}
\caption {\small{HPX Stream benchmark's strong scaling with/without using the proposed smart executors.}}
\label{f9}
\end{center}
\end{figure}

This benchmark \cite{str1, str2} has been widely used for evaluating memory bandwidth of a system. In \cite{hpx3}, the HPX executors performance were evaluated on this benchmark with $50$ million data points over $10$ iterations. The other characteristic information of this loop is included in Table \ref{t3}. As shown in Fig.\ref{f14}, this application includes $4$ operations over $3$ equally sizes arrays ($A$, $B$ and $C$) that are: copy ($C = A$), scale ($B = k \times C$), adding ($C = A + B$) and triad ($A = B + K \times C$). All three proposed smart executors are applied on this loop to make HPX to choose an execution policy, chunk size and prefetching distance efficiently. The speedup comparison results of the data transform measurements with/without using proposed techniques are illustrated in Fig.\ref{f9}. As we can see, using the proposed smart executors together on this benchmark improves HPX performance by an average of about $13\%$ compared to using HPX auto-parallelization techniques without considering static/dynamic information and implementing machine learning technique.

\subsubsection{Stencil Benchmark}

\begin{figure}
\begin{center}
\centering
\includegraphics[width=0.7\columnwidth]{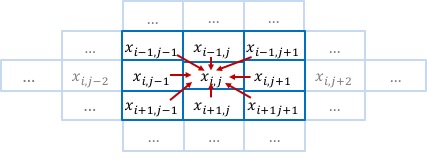}\vspace{-0.15cm}
\caption {\small{Heat Distribution Benchmark, $2D$ Stencil.}}
\label{f10}
\end{center}
\end{figure}

\begin{figure}
\begin{center}
\centering
\includegraphics[width=0.6\columnwidth]{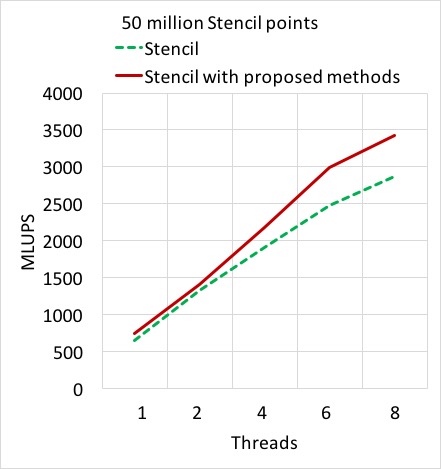}\vspace{-0.15cm}
\caption {\small{HPX $2D$ Stencil benchmark's strong scaling with/without using the proposed smart executors.}}
\label{f11}
\end{center}
\end{figure}

The performance of different HPX scheduling policies on a $2D$ Stencil benchmark is studied in \cite{hpx4}. This application is a two dimensional heat distribution shown in Fig.\ref{f10}, in which the temperature of each point is computed based on the temperature of its neighbors. The characteristic information of this loop is included in Table \ref{t3}. 
The speedup comparison results of HPX performance with/without using the proposed smart executors are illustrated in Fig.\ref{f11}. It shows HPX performance improvement by an average of about $22\%$ by using the proposed techniques together on this loop compared to using HPX auto-parallelization techniques without considering static/dynamic information and implementing machine learning technique.

\section{Conclusion and Future Works}
\label{sec:con}

The main goal of this paper is to illustrate a powerful new set of techniques that can be made available to application developers when compilers, runtime systems, and machine learning algorithms work in concert. These techniques developed here not only greatly improve performance, but users are able to reap their benefit with little cost to themselves. Simply by annotating their code with high level executors, users can see their application's performance increase in a portable way.

These results could have broad impact for applications and libraries as well as the maintainers and scientist that use them. The high level annotations increase the usability and therefore accessibility of runtime features that before would have taken a knowledgeable developer to implement. Due to the machine learning element, users will not have to worry about losing performance in different runtime environments that could manifest themselves. Finally, the inclusion of compiler information will allow these performance optimizations to be platform independent. These three features taken together present a notable solution to the challenges presented by an increasingly multi-core and heterogeneous world.

As powerful as these techniques may be, more work is needed to be done in order to fully realize the potential of this work. Notably, the breadth of performance characteristics needs to be more carefully studied to understand the core features that relate to performance. Additionally more research is needed to ensure that the characteristics measured here also are relevant for other architectures such as the new Knights Landing chipset. On a shorter time scale we intend to investigate extending the number of features for improving the resulting loop's parameters prediction.

In this paper, we have illustrated that the parallel performance of our test cases were improved by using a machine learning algorithm to determine either an appropriate code path (parallel or sequential loop execution) or certain parameters for the loop execution itself (chunk size or prefetching distance). The speedup results of these test cases and benchmarks showed by around $12\%-35\%$ improvement compared to selecting execution policy, chunk size and prefetching distance of a loop without using static/dynamic information and machine learning technique. These results proved that combining machine learning techniques, compiler information, and runtime methods helps an application maximize the available resources\footnotemark[1].

\footnotetext[1]{This works was supported by NSF awards 1447831 and 1339782.}

\bibliographystyle{unsrtnat}
\bibliography{References} 

\end{document}